\documentclass[aps,prc,preprint,superscriptaddress,showpacs]{revtex4}

\usepackage{graphicx}
\usepackage{textcomp}
\usepackage{amsfonts}
\usepackage{amssymb}
\usepackage{bm}
\usepackage[dvips]{color}

\begin{document}

\title{Sea flavor content of octet baryons and intrinsic five-quark Fock states}

\author{C. S. An}
\email[]{ancs@ihep.ac.cn}
\affiliation{Institut de Recherche sur les
lois Fondamentales de l'Univers, DSM/Irfu, CEA/Saclay, F-91191 Gif-sur-Yvette, France}

\affiliation{Institute of High Energy Physics, P. O. Box 918-4, Beijing 100049, China}

\affiliation{Theoretical Center for Science Facilities, CAS, P. O. Box 918-4, Beijing 100049, China}

\author{B. Saghai}
\email[]{bijan.saghai@cea.fr}
\affiliation{Institut de Recherche sur les lois Fondamentales de l'Univers, DSM/Irfu, 
CEA/Saclay, F-91191 Gif-sur-Yvette, France}

\date{\today}

\begin{abstract}

Sea quark contents of the octet baryons are investigated by employing
an extended chiral constituent quark approach, which embodies higher Fock five-quark components
in the baryons wave-functions. 
The well-known flavor asymmetry of the nucleon sea $\bar{d}-\bar{u}$, 
is used as input to predict the probabilities of $\bar{u}$, $\bar{d}$ and 
$\bar{s}$ in the nucleon, $\Lambda$, $\Sigma$ and $\Xi$ baryons, 
due to the intrinsic five-quark components in the baryons wave functions.

\end{abstract}

\pacs{12.39.-x, 14.20.-c, 14.65.-q, 14.65.Bt}
\maketitle

\section{Introduction}
\label{intro}
%
%
Though the baryons valence quark distributions are known to be flavor asymmetric, till mid-nineties
those of the sea quarks were assumed to be symmetric.
However, In early 80's, Thomas~\cite{Thomas:1983fh} predicted that the pion cloud dressing the proton can generate 
an enhancement of the light antiquarks flavor asymmetry, $\bar{d}-\bar{u}$.
The experimental benchmark in the flavor sea content of the proton appeared about one decade later due to the pioneer measurements 
by the New Muon Collaboration, which showed~\cite{Arneodo:1994sh} a significant excess of the anti-down relative 
to anti-up quark in the proton sea, as a function of the parton momentum fraction (Bjorken-x)
\begin{equation}
\bar{d}-\bar{u} = \int_{0}^{1} \Big [\bar{d}_p(x)-\bar{u}_p(x) \Big ] dx = 0.147 \pm 0.039\,.
\label{Bjo}
\end{equation} 
This unexpected large asymmetry between the down and up antiquarks distribution in the nucleon was confirmed by other measurements 
in various $0 \leq x \leq 1$ ranges at CERN~\cite{Baldit:1994jk}, 
Fermilab~\cite{Hawker:1998ty,Peng:1998pa,Towell:2001nh} and DESY~\cite{Ackerstaff:1998sr}.

Those measurements imply the breaking of the so-called Gottfried sum rule~\cite{Gottfried:1967kk}, which expressed in 
terms of parton distribution~\cite{Field:1976ve}, has the following form:
\begin{equation}
\mathcal{I}_G = \int_{0}^{1} {\Big [F_{2}^{p}(x) - F_{2}^{n}(x)\Big ]} \frac {dx}{x}= \frac {1}{3}
                - \frac {2}{3} \int_{0}^{1} \Big [\bar{d}_p(x)-\bar{u}_p(x)\Big ] dx, 
\label{Got}
\end{equation} 
where $F_{2}^{p}$ and $F_{2}^{n}$ are the proton and the neutron structure functions. A symmetric sea assumption gives 
the Gottfried sum rule: $\mathcal{I}_G = \frac {1}{3}$. (Note that the original sum rule published by 
Gottfried~\cite{Gottfried:1967kk} was much simpler and rather naive; for steps having led to the above expression
see, e.g. Sec. 2.2 in Ref.~\cite{Garvey:2001yq}.)  

Evidence for the breaking of the Gottfried sum rule motivated a large amount of effort 
to understand the origins of the nucleon sea, either perturbative or nonprturbative, as reviewed by several 
authors~\cite{Garvey:2001yq,Speth:1996pz,Kumano:1997cy,Vogt:2000sk,Peng:2003zm}.

Perturbative mechanisms are generated by the gluons splitting into quark-antiquark pairs.
The only nonperturbative gluonic process is due to gluon condensate~\cite{Diakonov:1995zi}, as investigated 
in the soliton based approaches. Other sources of the nonperturbative mechanisms are being extensively studied, 
as summarized in the following.
 
In the meson cloud scheme, a variety of approaches has been developed; namely,  
bag model~\cite{Thomas:1983fh,Henley:1990kw,Alberg:2012wr},
light-cone meson-baryon fluctuations of intrinsic $q \bar q$ pairs~\cite{Brodsky:1996hc,Cao:1999da,Cao:1999jc}, 
one-pion-exchange~\cite{Nikolaev:1998se}, 
meson-baryon effective Lagrangian~\cite{Carvalho:1999he,Huang:2007dy,Chen:2009xy}
and quantum fluctuations of the baryon~\cite{Alwall:2005xd}.
The light antiquarks are generated by the nucleon fluctuations into Fock states 
$|\pi N\rangle$ and/or $|\pi \Delta\rangle$. Then, $\bar{u}$ arises from $|\pi^- \Delta^{++}\rangle$, $\bar{d}$ 
from $|\pi^+ n\rangle$ and $|\pi^+ \Delta^\circ\rangle$, while $u\bar{u}$ and $d\bar{d}$ related to $\pi^\circ$ 
are assumed to annihilate. 
A rather comprehensive set of fluctuations includes the following Fock states~\cite{Huang:2007dy}: 
$|\pi N\rangle$, $|\rho N\rangle$, $|\omega N\rangle$, $|\pi \Delta\rangle$, $|\rho \Delta\rangle$,
$|K \Lambda\rangle$, $|K^* \Lambda\rangle$, $|K \Sigma\rangle$, $|K^* \Sigma\rangle$.
Finally, reggeizing~\cite{Nikolaev:1998se} the virtual mesons was a significant step in the meson cloud approaches.  

A more evolved meson cloud formulation is based on the large-$N_c$ limit of QCD, where the baryons are treated as
chiral solitons {\it via} collective excitations of 
mesons~\cite{Brodsky:1988ip,Diakonov:1997mm,Wakamatsu:1997en,Dressler:1998zi,Pobylitsa:1998tk,615110}. 

In the same line, chiral constituent quark models 
($\chi CQM$)~\cite{Huang:2007dy,Dahiya:2003ms,Shu:2007xb,885054,arXiv:1002.4747,Sharma:2010sm,Dahiya:2010ht,arXiv:1101.3378,Yuan:2012wz}
concentrate on the meson cloud, where the virtual pion couples directly to a quark. 
It is worth pointing out that within a $\chi CQM$~\cite{Sharma:2010sm} the breaking of the Gottfried sum rule also 
for baryons other than the nucleon is predicted.

Moreover, dedicated studies on the quark-antiquark decomposition in the sea quark are also being
extensively performed within effective QCD~\cite{Brodsky:1980pb,Brodsky:1981se,arXiv:1102.5631,arXiv:1105.2381},
lattice QCD~\cite{Young:2009ps,Takeda:2010cw,Bali:2011rs} and
statistical balance~\cite{Trevisan:2008zz,Zhang:2010ac}.

A pertinent non-perturbative source is due to genuine higher Fock components in the baryon wave function.
In 1981, in order to interpret the large cross section of charmed particle production in hadron collisions,
Brodsky and collaborators~\cite{Brodsky:1980pb,Brodsky:1981se} postulated the existence of the $|uudc\bar c\rangle$
configuration in the proton; called the $BHPS$ model. 
That approach was recently extended~\cite{arXiv:1102.5631,arXiv:1105.2381} 
to the light quarks sector, describing nicely data for $\bar{d}-\bar{u}$ and $\bar{u}+\bar{d}-s-\bar{s}$.
 
The present work is dedicated to the investigation of the flavor sea components, arising from the
five-quark components, in the ground state baryons; namely, $N$, $\Lambda$, $\Sigma$ and $\Xi$.

Our formalism is based on an extended chiral constituent quark approach and embodies all possible 
five-quark mixtures in the baryons wave-functions.
Such higher Fock-components have been proven to be quite significant in describing the 
properties of baryons, their electromagnetic and strong decays ~\cite{Zou:2005xy,Li:2005jn,An:2005cj,JuliaDiaz:2006av,Li:2005jb,Li:2006nm,An:2008xk,An:2008tz,An:2010wb,Zou:2009zz,Zou:2010tc,An:2011sb}.

In order to fix the only adjustable parameter of our model, we use as input the result extracted from the measurement performed by the 
FNAL E866/NuSea Collaboration~\cite{Towell:2001nh}
\begin{equation}
\bar{d}-\bar{u} = 0.118 \pm 0.012\,.
\label{E866}
\end{equation}

Using our model, we put forward predictions on the probabilities of $u \bar u$, $d \bar d$ and $s \bar s$.
Comparisons with results from other works are also reported.

The present manuscript is organized in the following way: in section~\ref{theo}, we present our theoretical 
formalism which includes the wave functions and couplings between three- and five-quark components.
Expressions for the couplings and the five-quark configurations energies are derived and all of the relevant 
associated orbital-flavor-spin configurations are singled out.
Numerical results are given in section~\ref{num}, putting forward predictions for the probabilities of different 
five-quark configurations, as well as those of the sea content of the baryons.
Finally section~\ref{sumcon} contains summary and conclusions.
%
%
\section{Theoretical Frame}
\label{theo}
In order to investigate the sea content of the octet baryons, we employ the extended constituent
quark model (E-$\chi CQM$), in which wave function for a baryon is expressed as
\begin{equation}
 |\psi\rangle_{B}=\frac{1}{\mathcal{\sqrt{N}}}\left[|QQQ\rangle+
 \sum_{i,n_{r},l}C_{in_{r}l}|QQQ(Q\bar{Q}),i,n_{r},l\rangle \right]\,,
\label{wfn}
\end{equation}
where the first term is the conventional wave function for the baryon with three
constituent quarks, and the second term is a sum over all possible higher Fock
components with a $Q\bar{Q}$ pair.
Here we denote light quark-antiquark pair as $Q\bar{Q}\equiv q\bar{q}$ (with $q \equiv u,~d,$)
and strange quark-antiquark pairs as $Q\bar{Q}\equiv s\bar{s}$.
Different possible orbital-flavor-spin-color configurations of the four-quark
subsystems in the five-quark system are numbered by $i$; $n_{r}$ and $l$ denote the inner radial and
orbital quantum numbers, respectively. 
Finally, $C_{in_{r}l}/\sqrt{\mathcal{N}}\equiv A_{in_{r}l}$
represents the probability amplitude for the corresponding five-quark component.

In the present case, we consider the ground states of baryon octet, whose parities
are positive, so that the orbital quantum number $l$ must be an odd number $1,3,\cdots 2n+1$.
The total spin $S$ of a five-quark system can only be $\frac{1}{2}$, $\frac{3}{2}$
or $\frac{5}{2}$, so $l$ cannot be higher than $3$ to combine with $S$,
forming spin $\frac{1}{2}$ for the baryons considered here. 
All of the five-quark configurations with $l=1$ and $n_r=0$, which may form higher Fock components in 
the proton~\cite{An:2005cj}, can directly be extended to other baryons of the octet. 
We will discuss later the five-quark configurations with $l=3$ and $n_r \neq 0$.

The coefficients $C_{in_{r}l}$ in Eq.~(\ref{wfn}) can be related to the 
coupling between the valence three-quark and the corresponding five-quark components
\begin{equation}
 C_{in_{r}l}=\frac{\langle QQQ(Q\bar{Q}),i,n_{r},l|\hat{T}|QQQ\rangle}{M_{B}-E_{in_{r}l}}\,,
\end{equation}
where $\hat{T}$ is a model dependent coupling operator, $M_{B}$ the mass of baryon B and $E_{in_{r}l}$
the energy of the five-quark component, as discussed in the next two Subsections. 

\subsection{Couplings between three- and five-quark components for baryon octet}
\label{cou}

Here we use a $^{3}P_{0}$ version for the transition coupling operator $\hat{T}$
\begin{equation}
 \hat{T}=-\gamma\sum_{j}\mathcal{F}_{j,5}^{00}\mathcal{C}_{j,5}^{00}C_{OFSC}
\left [\sum_{m}
\langle1,m;1,-m|00\rangle\chi^{1,m}_{j,5}
\mathcal{Y}^{1,-m}_{j,5}(\vec{p}_{j}-\vec{p}_{5})b^{\dag}(\vec{p}_{j})d^{\dag}(\vec{p}_{5}) \right ]\,,
\label{op}
\end{equation}
where $\gamma$ is a dimensionless constant of the model as discussed later, 
$\mathcal{F}_{i,5}^{00}$ and $\mathcal{C}_{i,5}^{00}$ 
denote the flavor and color singlet of the quark-antiquark pair $Q_{i}\bar{Q}$ in the five-quark 
system, and $C_{OFSC}$ is an operator to calculate the orbital-flavor-spin-color overlap between 
the residual three-quark configuration in the five-quark system and the valence three quark system.

To derive the matrix elements of $\hat{T}$ between the three- and five-quark configurations,
we need explicit wave functions for the latter ones. 
As shown in Ref.~\cite{An:2005cj}, if we limit the orbital quantum and radial quantum numbers to 
$l=1$ and $n_{r}=0$, respectively, then, there are 24 different five-quark configurations which can form 
possible components in the proton. 
For the four-quark subsystem in these five-quark configurations, the orbital wave functions are 
$[4]_{X}$ or $[31]_{X}$, flavor wave functions $[31]_{F}$, $[22]_{F}$ or $[211]_{F}$ and spin wave 
functions $[4]_{S}$, $[31]_{S}$ or $[22]_{S}$. 
Explicit forms of these wave functions for the proton can be found in Refs.~\cite{An:2005cj,chen}
and using the same approach allows inferring the wave functions for the other octet baryons.
Notice that there are two separate classes of wave functions with the flavor 
symmetry $[31]_{F}$, relation and difference between these two are explained in Ref.~\cite{An:2005cj}. 
Here we denote these two as $[31]_{F}^{1}$ and $[31]_{F}^{2}$.
So the total number of five-quark configurations goes up frpm 24 to 34.
The $SU(2)$ Clebsch-Gordan coefficients (C-G) for the combinations of the four classes of four-quark 
flavor configurations with an antiquark to form the required four categories of 
isospin eigenstates $N$, $\Lambda$, $\Sigma$ and $\Xi$ are listed in Table~\ref{cg}.

As discussed above, we have to consider the five-quark configurations with $l=3$. 
In this case, total spin $S$ of the five-quark system should be $\frac{5}{2}$, 
so that the spin
wave function of the four-quark subsystem in such five-quark configurations must be $[4]_{S}$,
namely completely symmetric. 
Our calculations show that the couplings of a three-quark system to that set of five-quark 
configurations vanish. 
Consequently, five-quark configurations with $l=3$ cannot exist in 
the ground state of baryon octet.

With respect to the five-quark configurations with $n_{r} \geq 1$, 
the probabilities of these excitations in the octet baryons turn out to be negligible, because 
on the one hand the matrix elements of the coupling transition operator $\hat{T}$ 
between three-quark and $n_{r}\geq1$ five-quark components are much smaller than those 
for $n_{r}=0$, and on the other hand the energies should be at least several hundreds MeV higher. 

Consequently, we consider only the five-quark configurations with $l=1,~n_{r}=0$
as candidates of higher Fock components in the octet baryons. 
The derived matrix elements $T$ for the 34 five-quark configurations show
that only 17 configurations survive and matrix elements $T$ for all other ones vanish. 
We list the former configurations in Table~\ref{con}. 
The results for $T$ are listed in Tables~\ref{eng1q}, \ref{eng1s} and~\ref{eng0}. 
Note that the full coupling matrix elements is obtained by multiplying each term listed in the Tables 
by a common factor $V$
\begin{equation}
V=\gamma\omega_{5}C_{35}\,,
\end{equation}
with
\begin{equation}
C_{35}\equiv \Big ( \frac{2\omega_{3}\omega_{5}}{\omega_{3}^{2}+\omega_{5}^{2}}\Big )^{3}\,,
\end{equation}
where $\omega_{3}$ and $\omega_{5}$ are the harmonic oscillator parameters for 
the three- and five-quark components in baryons. 
The parameter $\omega_{3}$ can be inferred from the empirical radius of the proton {\it via}
$\omega_{3}=1/\sqrt{\langle r^{2}\rangle}$, which yields $\omega_{3}\simeq 246$ MeV if we 
take $\sqrt{\langle r^{2}\rangle}=1$~fm. 
Moreover, if the confining potential for the quarks is taken to be color dependent~\cite{Glozman:1995fu}, 
we can simply obtain the relation between $\omega_{3}$ and $\omega_{5}$ as
\begin{equation}
 \omega_{5}=\sqrt{\frac{5}{6}}\omega_{3}\,,
 \label{omega5}
\end{equation}
which leads to $\omega_{5}\simeq 225$~MeV.  
%
%
\subsection{Energies of five-quark components}
\label{eng}
The five-quark configurations listed in Tables~\ref{eng1q}, \ref{eng1s} and~\ref{eng0} share
the same energy if we neglect the hyperfine interaction between the quarks
and the constituent mass difference between the light and strange quarks;
we denote this energy as $E_{0}$. 
Then, the energy of a given five-quark configuration with number $i$ reads
\begin{equation}
 E_{i}=E_{0}+\Delta E_{i},
\label{E}
\end{equation}
where,
\begin{equation}
\Delta E_{i}\equiv E_{i}^{h}+n_{i}^{s}\delta m\,,
\label{dE}
\end{equation}
with $E_{i}^{h}$ the energy caused by hyperfine interaction 
between quarks, $n_{i}^{s}$ the number of strange quarks in the corresponding five-quark system 
and $\delta m = m_{s}-m$ the mass difference between the constituent strange and light quarks. 
To consider the hyperfine interaction between quarks, we employ the flavor-spin dependent version 
in the chiral constituent quark model~\cite{Glozman:1995fu},
\begin{equation}
 H_{h}=-\sum_{i<j}\vec{\sigma}_{i}\cdot\vec{\sigma}_{j}
                    \left [ \sum_{a=1}^{3}V_{\pi}(r_{ij})\lambda^{a}_{i}\lambda^{a}_{j}+
                   \sum_{a=4}^{7}V_{K}(r_{ij})\lambda^{a}_{i}\lambda^{a}_{j}+
                   V_{\eta}(r_{ij})\lambda^{8}_{i}\lambda^{8}_{j} \right ]\,,
\label{hyp}
\end{equation}
where $\lambda^{a}_{i}$ denotes the Gell-Mann matrix acting on the $i^{th}$ quark, $V_{M}(r_{ij})$
is the potential of the $M$ meson-exchange interaction between $i^{th}$ and $j^{th}$ quark. Since
the hyperfine interaction between a quark and an antiquark is negligible, 
after taking into account the overall symmetry of the wave functions, $H_{h}$ can simply be
replaced by an operator acting on the first two quarks 
\begin{equation}
H_{h}=-6\vec{\sigma}_{1}\cdot\vec{\sigma}_{2}
                    \left [ \sum_{a=1}^{3}V_{\pi}(r_{12})\lambda^{a}_{1}\lambda^{a}_{2}+
                   \sum_{a=4}^{7}V_{K}(r_{12})\lambda^{a}_{1}\lambda^{a}_{2}+
                   V_{\eta}(r_{12})\lambda^{8}_{1}\lambda^{8}_{2} \right ]\,.
\label{shyp}
\end{equation}
Then, $\Delta E_i$ in terms of $E_{i}^{h}$, Eq.~(\ref{dE}), is obtained by
\begin{eqnarray}
E^{h}_{i}&=&\langle QQQ(Q\bar{Q}),i,0,1|H_{h}|QQQ(Q\bar{Q}),i,0,1\rangle  \nonumber\\
&=&-6 \sum_{njklm} 
\Bigg [ 
(C_{[31]^{n}_{i}[211]_{n}}^{[1^4]})^{2}C^{[31]_{i}^{n}}_{[\mathcal{FS}]^{j}_{i}
[\mathcal{X}]_{i}^{l}}
C^{[31]_{i}^{n}}_{[\mathcal{FS}]^{k}_{i}[\mathcal{X}]_{i}^{m}}
\nonumber\\
&& 
\Big (
\langle[\mathcal{X}]^{l}_{i}|V_{\pi}(r_{12})|[\mathcal{X}]^{m}_{i}\rangle
\langle[\mathcal{FS}]^{j}_{i}|\vec{\sigma}_{1}\cdot\vec{\sigma}_{2}\sum_{a=1}^{3}\lambda^{a}_{1}
\lambda^{a}_{2}|[\mathcal{FS}]^{k}_{i}\rangle
\nonumber\\
&& +\langle[\mathcal{X}]^{l}_{i}|V_{K}(r_{12})|[\mathcal{X}]^{m}_{i}\rangle
\langle[\mathcal{FS}]^{j}_{i}|\vec{\sigma}_{1}\cdot\vec{\sigma}_{2}\sum_{a=4}^{7}\lambda^{a}_{1}
\lambda^{a}_{2}|[\mathcal{FS}]^{k}_{i}\rangle
\nonumber\\
&& +\langle[\mathcal{X}]^{l}_{i}|V_{\eta}(r_{12})|[\mathcal{X}]^{m}_{i}\rangle
\langle[\mathcal{FS}]^{j}_{i}|\vec{\sigma}_{1}\cdot\vec{\sigma}_{2}
\lambda^{8}_{1} \lambda^{8}_{2}|[\mathcal{FS}]^{k}_{i}\rangle \Big ) 
\Bigg ]
\,,
\end{eqnarray}
where $[\mathcal{FS}]^{\it N}_{i}$ and $[\mathcal{X}]^{\it N}_{i}$ represent
the ${\it N}^{th}$ flavor-spin and orbital wave functions
of the four-quark subsystem in the five-quark configuration with number $i$
of the 17 five-quark configurations listed in Table~\ref{con}, respectively.
$C_{[31]^{n}_{i}[211]^{n}}^{[1^4]}$, $C^{[31]_{i}^{n}}_{[\mathcal{FS}]^{j}_{i}[\mathcal{X}]_{i}^{l}}$
and $C^{[31]_{i}^{n}}_{[\mathcal{FS}]^{k}_{i}[\mathcal{X}]^{i}_{m}}$ are the $S_{4}$ Clebsch-Gordan coefficients.

As discussed in Section~\ref{cou}, we need to consider only the five-quark configurations
with the spin of the four-quark subsystem being $[22]_{S}$ and $[31]_{S}$. 
Explicit calculations lead to the following matrix elements:
\begin{eqnarray}
\langle[22]_{S1}|\vec{\sigma}_{1}\cdot\vec{\sigma}_{2}|[22]_{S1}\rangle&=&1,\\
\langle[22]_{S2}|\vec{\sigma}_{1}\cdot\vec{\sigma}_{2}|[22]_{S2}\rangle&=&-3\,,\\
\langle[31]_{S1}|\vec{\sigma}_{1}\cdot\vec{\sigma}_{2}|[31]_{S1}\rangle&=&1,\\
\langle[31]_{S2}|\vec{\sigma}_{1}\cdot\vec{\sigma}_{2}|[31]_{S2}\rangle&=&1,\\
\langle[31]_{S3}|\vec{\sigma}_{1}\cdot\vec{\sigma}_{2}|[31]_{S3}\rangle&=&-3\,.
\end{eqnarray}

Now we have to consider the matrix elements of the flavor operators, which are
linear combinations of the spatial matrix elements of the two-body potential 
$V_{M}(r_{12}),~M \equiv \pi,K,\eta$, which are defined as
\begin{equation}
P_{l}^{M}=\langle lm(r_{12})|V_{M}(r_{12})|lm(r_{12})\rangle\,, 
\end{equation}
where $|lm\rangle$ represents the spatial wave function.
Within exact flavor $SU(3)$ symmetry, $P^{\pi}_{l}=P_{l}^{K}=P^{\eta}_{l}$ and explicit calculations 
for matrix elements of the flavor-dependent operator lead to the hyperfine interaction energies given 
in ref.~\cite{An:2005cj}. 
To analyze the flavor asymmetry of the sea contents, including $\bar{u}$, $\bar{d}$ and $\bar{s}$, in the baryon octet, 
we have to take into account the flavor $SU(3)$ breaking, which implies $P^{\pi}_{l}\neq P_{l}^{K}\neq P^{\eta}_{l}$.
In addition, we have to treat properly the three subsets of the sea components; namely, the $\eta$-exchange interaction 
for pairs of light quarks($V_{u\bar{u}}(r_{12})$~or~$V_{d\bar{d}}(r_{12})$), 
one light and one strange ($V_{u\bar{s}}(12)$~or~$V_{d\bar{s}}(r_{12})$) and 
two strange quarks ($V_{s\bar{s}}(r_{12})$). 
We take $P^{u\bar{u}}_{l}=P^{d\bar{d}}_{l}=P^{\pi}_{l}$ and $P^{u\bar{s}}_{l}=P^{d\bar{s}}_{l}=P^{K}_{l}$. 
The empirical values for $P^{M}_{l}$ with $l=0,1$ are taken from Ref.~\cite{Glozman:1995fu}
\begin{eqnarray}
& P^{\pi}_{0}=29~MeV,~P^{K}_{0}=20~MeV,~P^{s\bar{s}}_{0}=14~MeV\,,\\
& P^{\pi}_{1}=45~MeV,~P^{K}_{1}=30~MeV,~P^{s\bar{s}}_{1}=20~MeV\,.
\end{eqnarray}
%
\subsection{Probabilities of sea quark components}
In Table~(\ref{nump}) our results for probabilities of $q \bar{q}$ ($q \equiv u, ~d$) and 
$s \bar{s}$ are given for each of the 17 five-quark configurations reported in 
Table~(\ref{con}). 
Using expressions in Tables~(\ref{eng1q}) to~(\ref{eng0}), we get the probabilty of the sea quark in
each baryon $B$
\begin{equation}
P_B^{Q\bar{Q}}= \frac{1}{\mathcal{N}}
\sum_{i=1}^{17}\Bigg[ \Big ( \frac{T_i^{Q\bar{Q}} }{M_B-E_i^{Q\bar{Q}}} \Big )^2 \Bigg ]\,.
\label{prob}
\end{equation}
where the normalization factor reads
\begin{eqnarray}
\mathcal{N} & \equiv & 1+ \sum_{i=1}^{17} \mathcal{N}_i \\  
&=& 1+\sum_{i=1}^{17} \Bigg[ \Big ( \frac{T_i^{u\bar{u}}}{M_B-E_i^{u\bar{u}}}\Big )^2 +
\Big (\frac{T_i^{d\bar{d}}}{M_B-E_i^{d\bar{d}}}\Big )^2 +
\Big ( \frac{T_i^{s\bar{s}}}{M_B-E_i^{s\bar{s}}} \Big )^2 \Bigg ].
\label{norm}
\end{eqnarray}
Notice that in Eq. (24) the first term is due to the valence three-quark states, while the second term comes 
from the five-quark mixtures.
%
\section{Numerical results and discussion}
\label{num}
%
\subsection{Adjustable parameters}
\label{aps}
In addition to the values given in Eqs.~(9,~21,~22), for quarks masses difference, 
we used the common value $\delta m\equiv m_s - m_q =120$ MeV.

To get the numerical results, we still need to determine the values for two other parameters, 
$E_{0}$ and $V$. 

The first parameter $E_0$, Eq.~(\ref{E}), is determined from other sources, as explained below. 
This quantity can be calculate employing a constituent quark model approach
\begin{equation}
E_{0}=\sum_{j=1}^{5}m_{j}+7\omega_{5}+5V_{0}\,,
\end{equation}
where $m_{j}$ denote the constituent quark mass and $V_{0}$ a model parameter 
which represents the energy contributed by the inharmonic part of the potential 
for the five-quark system. 
Consequently, $E_{0}$ is dependent on three parameters. 
The value for $V_{0}$ not being known well enough, we use themeson cloud approach to 
fix directly $E_{0}$.
Actually, the first two five-quark configurations listed in Table~\ref{con} 
are very similar to the $\pi N$ and $\pi \Delta$ meson clouds, and a commonly 
accepted~\cite{Garvey:2001yq} ratio for the probabilities of the 
former to the latter one in the nucleon is $2$. 
Here we use this value to determine the probabilities of the various five-quark 
configurations for the octet baryons, which leads to
\begin{equation}
 E_{0} = 2127~MeV.
\end{equation}

The only adjustable parameter of our model is then $V$, determined using the data. 
The flavor asymmetry $\bar{d}-\bar{u}$ of the proton, related to the probabilities of $q \bar{q}$ 
components $\mathcal{P}_{5q}^{q \bar{q}}$, is given by the following expression:
\begin{eqnarray}
\mathcal{P}_{5q}^{d \bar{d}}- \mathcal{P}_{5q}^{u \bar{u}}&\equiv& \bar{d}-\bar{u}\nonumber\\ 
&=& \frac{1}{\mathcal{N}}
\Bigg\{ \Big ( \frac{T_{1}^{q \bar{q}}}{M_p-E_1}\Big )^2 + 
\Big ( \frac{T_6^{q \bar{q}}}{M_p-E_6}\Big )^2 + 
\Big ( \frac{T_{13}^{q \bar{q}}}{M_p-E_{13}}\Big )^2\nonumber\\
&-& \frac{1}{3} \bigg[ \Big (\frac{T_2^{q \bar{q}}}{M_p-E_2}\Big )^2 + 
\Big (\frac{T_7^{q \bar{q}}}{M_p-E_7})\Big )^2 +
\Big (\frac{T_9^{q \bar{q}}}{M_p-E_9}\Big )^2 \nonumber\\
&+& \Big ( \frac{T_{14}^{q \bar{q}}}{M_p-E_{14}}\Big )^2 + 
\Big ( \frac{T_{16}^{q \bar{q}}}{M_p-E_{16}} \Big )^2\bigg] \Bigg\}\,,
\label{Asym}
\end{eqnarray}
$T_i^{q \bar{q}}$ are linear functions of $V(\bar{u}, \bar{d})$, the value of which is adjusted by 
using as input the data~\cite{Towell:2001nh} for the flavor asymmetry $\bar{d}-\bar{u}$ of the nucleon
\begin{equation}
\bar{d}-\bar{u} = 0.118 \pm 0.012,
\end{equation}
leading to
\begin{equation}
 V(\bar{u}, \bar{d}) = 570 §\pm§ 46~MeV.
\end{equation}
\subsection{Results and discussion}
\label{redi}
In Table~(\ref{nump}) our results for probabilities of five-quark components, arising from the 17 
configaurations given in Table~(\ref{con}), are reported for the studied baryons.

The configuration $[31]_{X}[4]_{FS}[22]_{F}[22]_{S}$ turns out to be the dominant one for
all of the considered ground state baryons.
The following most important ones are 
$[31]_{X}[4]_{FS}[31]^{2}_{F}[31]_{S}$ for $\Lambda$
and $[31]_{X}[4]_{FS}[31]^{1}_{F}[31]_{S}$ for the other three baryons; 
$[4]_{X}[31]_{FS}[22]_{F}[31]_{S}$ plays also a significant role for $\Xi$.
Finally, the other major configurations are 
$[4]_{X}[31]_{FS}[31]^{2}_{F}[22]_{S}$ for $\Lambda$ and
$[4]_{X}[31]_{FS}[31]^{1}_{F}[22]_{S}$ for the three other baryons. 
Added up Contributions from those configurations, embody the $q \bar{q}$ components at the level of 83\% for $N$, 72\% for 
$\Lambda$ and about 65\% for $\Sigma$ and $\Xi$.
In the case of $s \bar{s}$ the contributions from different configurations show much less 
variations than in the case of $q \bar{q}$.

Predictions of our model for the sea content of the octet baryons, in particle basis, are given 
in Table~(\ref{sea}), with, in addition, extracted values for $\bar{d}-\bar{u}$, $\bar{u}+\bar{d}$ and 
$\bar{u}+\bar{d}+\bar{s}$.

Our result for the total sea probability in the proton, Table~(\ref{sea}) $7^{th}$ column, 
is close to those reported by Chang and Peng~\cite{arXiv:1105.2381}. 
This latter work is a generalization of the approach developed by 
Brodsky and collaborators~\cite{Brodsky:1980pb,Brodsky:1981se} (G-BHPS)
investigating $uudc \bar{c}$ five-quark components in the proton. 
The most significant difference between our results and those in Ref.~\cite{arXiv:1105.2381} 
concerns the $s \bar{s}$ component.

For $u \bar{u}$ and $d \bar{d}$ components of the sea in all baryons studied here,
predictions have been reported by Shao and collaborators~\cite{arXiv:1002.4747}, both in
chiral quark ($\chi$CQM) and meson cloud (MC) approaches. 
The chiral quark results~\cite{arXiv:1002.4747} show significant discrepancies with
our findings. The most drastic case concerns the proton and only results for $\Xi^\circ$
agree in the two approaches within 2$\sigma$. 
Notice that in Ref.~\cite{arXiv:1002.4747} effects arising from the $SU(3)$ symmetry
breaking have not been included. 
Results coming from the meson cloud span a rather large ranges which include our results, 
except in the case of $\Sigma^\circ$ and, to a lesser extent, $\Xi^\circ$. 
In Ref.~\cite{arXiv:1002.4747} the same probability has been introduced for $\pi N$, $\pi \Lambda$ and $\pi \Sigma$,
which is an ad hoc assumption.

Within a QCD-inspired unquenched quark model (UqQM) flavor asymmetries for ground state baryons
were investigated~\cite{885054}. Compared to our results, the main trend of the UqQM predictions show
an overestimate. However the relative flavor asymmetry  $A_{\Xi^\circ}/A_{p}$ turns out to
be by far much closer within UqQM and our results as compared with those reported by 
Shao and collaborators~\cite{arXiv:1002.4747}.

In the case of $\Sigma^+$ sea, we find $\bar{d} > \bar{u} > \bar{s}$, as in the case of the proton.
Our results endorse findings in meson cloud sector, e.g. within light-cone meson-baryon
fluctuations~\cite{Cao:1999jc} and meson-baryon effective Lagrangian~\cite{Carvalho:1999he}
approaches.

Another relevant quantity is the suppression factor ($\kappa$) of the nucleon strange quark content 
with respect to the non-strange sea quarks
\begin{equation}
\kappa = \frac{\int_{0} ^{1} \left [x s(x) + x\bar{s}(x) \right ] dx}{\int_{0}^{1} \left [x\bar{u}(x) + x\bar{d}(x)\right ] dx }
\approx \frac{2P_{s \bar{s}}}{ P_{u \bar{u}} + P_{d \bar{d}}},
\label{kappa}
\end{equation}
where $\kappa =1$ would indicate a flavor $SU(3)$ symmetric sea, while 
the CCFR collaboration~\cite{Bazarko:1994tt} has reported $\kappa$=0.48$\pm$0.05 
(see also Ref.~\cite{Conrad:1997ne}).
The present work leads to $\kappa$=0.4, in good enough agreement with the data.
In Ref.~\cite{arXiv:0710.5032}, investigating the "NuTeV anomaly"~\cite{Zeller:2001hh} within a penta-quark
model~\cite{Zou:2005xy}, that quantity was fixed at $\kappa \approx$0.5, and provided $P_{s \bar s}$=(3-20)\%.
The upper limit of $P_{s \bar s}$ allows explaining about 10\% of the anomaly within that approach.
Other possible sources to partially explain the "NuTeV anomaly" can be found in Ref.~\cite{Londergan:2007zza}.
Notice that the present work leads to $P_{s \bar s}$=(5.7$\pm$0.6)\%.
%
%
\section{Summary and Conclusions}
\label{sumcon}

An extended chiral constituent quark model, embodying genuine five-quark mixture in the ground state baryon octet
wave functions, was presented, focusing on the sea quark content.
 
The formalism leads to a model with only one adjustable parameter, the value of which is fixed using the measured
flavor excess of $\bar d$ over $\bar u$ in the proton.

We examined all possible 34 five-quark configurations in baryon octet and showed that only 17 of them, corresponding
to the orbital quantum number $l=1$ and radial quantum number $n_r=0$, are relevant to the higher Fock components in
the ground state octet baryons. 
Our formalism allows determining contributions from each of the 17 orbital-flavor-spin configurations and identifying the most 
significant ones for each baryon. One onfiguration, $[31]_{X}[4]_{FS}[22]_{F}[22]_{S}$, comes out as the dominant one for all 
the four investigated baryons. Five other configurations play major roles in one or another baryon. 
We put forward predictions of our complete model for the percentage per flavor, of the sea quark contents for $N$, $\Lambda$, $\Sigma$ and $\Xi$.
Finally, our predictions showed that the five-quark mixture in those baryons is around 30\%, of which about one fifth is due to the strange quark.

Better understanding of this nonperturbative mechanism is of course of paramount importance in the hadron spectroscopy 
and description of properties of baryons. 
But, this realm has also a crucial role in other issues related to foreseen measurements,
e.g. using electromagnetic probes, proton-proton collisions, neutrino scattering, WIMP search, as outlined below.

A comprehensive tomography of the nucleon is a part of the physics program~\cite{Avakian:2012ca} of the Jefferson
Laboratory (JLab) 12 GeV upgrade.

Strange sea content of the nucleon is also an important component in the processes in high-energy hadron colliders, 
such as W production mechanisms investigated at the LHC. 
A recent work~\cite{Yang:2009hh} shows that the $W$-boson production at the RHIC and LHC proton-proton colliders, 
would provide a unique opportunity in extracting the $\bar d / \bar u$ flavor asymmetry in the proton. 
Another recent investigation~\cite{Brodsky:2012vg} emphasizes that finding, and access to rich 
information on the intrinsic sea quark content of the proton, within the fixed-target experiment(AFTER) thanks to 
the LHC beam. 

Using a Tevatron-based neutrino beam, the high energy neutrino scattering experiment~\cite{Adams:2009kp}, 
Neutrino Scattering On Glass (NuSOnG), can allow measuring
the strange sea in the nucleon through charged current opposite sign dimuon production {\it via} the following 
two step reactions:
\begin{eqnarray}
\nu _\mu + N  &\to& \mu^- + c + X~;~c \to s + \mu^+ + \nu _\mu, \\
\bar{\nu}_\mu + N & \to& \mu^+ + \bar{c} + X~;~\bar{c} \to \bar{s} + \mu^- + \bar{\nu} _\mu,
\label{dimu}
\end{eqnarray}
improving significantly the data accuracy compared to that released by the NuTeV Collaboration~\cite{Zeller:2001hh}. 

Moreover, as emphasized in Ref.~\cite{Dinter:2011zz}, the strange content of the nucleon is an important ingredient 
in the dark matter search. Actually, the WIMP coupling to the nucleon would proceed through 
coupling of the Higgs boson to the scalar quark content of the nucleon. The dark matter cross section has
been found dominated by the strange quark content of the proton, see Ref.~\cite{Giedt:2009mr} and references therein.

Future measurements will allow us deepening our understanding, of both perturbative and nonperturbative mechanisms, 
on the origins of antiquarks in the baryons, and their intrinsic sea quark contents.

%
\begin{acknowledgments}

C.~S.~A. is supported by the European Commission within the CEA-Eurotalents Program.

\end{acknowledgments}
%

\newpage

%
%

\newpage
{\squeezetable
\begin{table}[h]
\caption{\footnotesize $SU(2)$ Clebsch-Gordan coefficients for the five-quark components with $q\bar{q}$.}

\begin{tabular}{lccccccccccccccccccccccccccccccc}
\hline \hline
             &&&&&&                &&&&&& $[31]_{F}^{1}$ &&&&&& $[31]_{F}^{2}$      &&&&&&   $[22]_{F}$ &&&&&& $[211]_{F}$\\
\hline

p            &&&&&&  $uudu\bar{u}$ &&&&&& $\sqrt{\frac{2}{3}}$  &&&&&&   $0$         &&&&&& $0$           &&&&&& $0$ \\
             &&&&&&  $uudd\bar{d}$ &&&&&& $\sqrt{\frac{1}{3}}$  &&&&&&   $0$            &&&&&& $1$           &&&&&& $0$ \\
n            &&&&&&  $uddu\bar{u}$ &&&&&& $\sqrt{\frac{1}{3}}$  &&&&&&    $0$          &&&&&& $-1$           &&&&&& $0$ \\
             &&&&&&  $uddd\bar{d}$ &&&&&& $\sqrt{\frac{2}{3}}$  &&&&&&    $0$          &&&&&& $0$           &&&&&& $0$ \\
$\Lambda$    &&&&&&  $udsu\bar{u}$ &&&&&& $0$                  &&&&&&$\sqrt{\frac{1}{2}}$ &&&&&& $-\sqrt{\frac{1}{2}}$ 
             &&&&&&  $-\sqrt{\frac{1}{2}}$  \\
             &&&&&&  $uusd\bar{d}$ &&&&&& $0$                  &&&&&&$\sqrt{\frac{1}{2}}$ &&&&&& $-\sqrt{\frac{1}{2}}$   
             &&&&&&   $-\sqrt{\frac{1}{2}}$  \\
$\Sigma^{+}$ &&&&&&  $uusu\bar{u}$ &&&&&&  $\sqrt{\frac{3}{4}}$ &&&&&&$0$   &&&&&& $0$           &&&&&& $0$ \\
             &&&&&&  $uusd\bar{d}$ &&&&&& $\sqrt{\frac{1}{4}}$  &&&&&& $1$  &&&&&& $1$           &&&&&& $-1$ \\
$\Sigma^{0}$ &&&&&&  $udsu\bar{u}$ &&&&&& $\sqrt{\frac{1}{2}}$ &&&&&&$\sqrt{\frac{1}{2}}$ &&&&&& $-\sqrt{\frac{1}{2}}$ 
             &&&&&&  $\sqrt{\frac{1}{2}}$  \\
             &&&&&&  $uusd\bar{d}$ &&&&&& $-\sqrt{\frac{1}{2}}$ &&&&&&$-\sqrt{\frac{1}{2}}$ &&&&&& $\sqrt{\frac{1}{2}}$   
             &&&&&&   $-\sqrt{\frac{1}{2}}$  \\
$\Sigma^{-}$ &&&&&&  $ddsu\bar{u}$ &&&&&& $\sqrt{\frac{1}{4}}$            &&&&&& $1$ &&&&&& $-1$           &&&&&& $1$ \\
             &&&&&&  $ddsd\bar{d}$ &&&&&& $\sqrt{\frac{3}{4}}$            &&&&&& $0$ &&&&&& $0$           &&&&&& $0$ \\
$\Xi^{0}$    &&&&&&  $ussu\bar{u}$ &&&&&& $\sqrt{\frac{2}{3}}$  &&&&&& $0$ &&&&&& $-\sqrt{\frac{2}{3}}$ 
             &&&&&&  $0$  \\
             &&&&&&  $ussd\bar{d}$ &&&&&& $\sqrt{\frac{1}{3}}$ &&&&&& $1$ &&&&&& $-\sqrt{\frac{1}{3}}$   
             &&&&&&   $-1$  \\
$\Xi^{-}$    &&&&&&  $dssu\bar{u}$ &&&&&& $\sqrt{\frac{1}{3}}$ &&&&&& $1$ &&&&&& $-\sqrt{\frac{1}{3}}$ 
             &&&&&&  $1$  \\
             &&&&&&  $dssd\bar{d}$ &&&&&& $\sqrt{\frac{2}{3}}$ &&&&&& $0$ &&&&&& $-\sqrt{\frac{2}{3}}$   
             &&&&&&   $0$  \\
\hline \hline
\end{tabular}
\label{cg}
\end{table}
}
%

{\squeezetable
\begin{table}[h]
\caption{Orbital-flavor-spin configurations for the five-quark states, relevant to the groud state
octet baryons.}
\begin{tabular}{cccccc}
\hline\hline
i & 1      &   2    &   3   &   4  &  5   \\

Config. & $[31]_{X}[4]_{FS}[22]_{F}[22]_{S}$ &  $[31]_{X}[4]_{FS}[31]^{1}_{F}[31]_{S}$ & $[31]_{X}[4]_{FS}[31]^{2}_{F}[31]_{S}$
              & $[31]_{X}[31]_{FS}[211]_{F}[22]_{S}$ & $[31]_{X}[31]_{FS}[211]_{F}[31]_{S}$  \\


i & 6     &  7    &     8    &   9   &  10 \\

Config. &   $[31]_{X}[31]_{FS}[22]_{F}[31]_{S}$ &   $[31]_{X}[31]_{FS}[31]^{1}_{F}[22]_{S}$ & $[31]_{X}[31]_{FS}[31]^{2}_{F}[22]_{S}$

    &  $[31]_{X}[31]_{FS}[31]^{1}_{F}[31]_{S}$ & $[31]_{X}[31]_{FS}[31]^{2}_{F}[31]_{S}$ \\

i & 11   &  12   &  13  &  14    &  15  \\

Config. &    $[4]_{X}[31]_{FS}[211]_{F}[22]_{S}$ & $[4]_{X}[31]_{FS}[211]_{F}[31]_{S}$ & 

        $[4]_{X}[31]_{FS}[22]_{F}[31]_{S}$ &   $[4]_{X}[31]_{FS}[31]^{1}_{F}[22]_{S}$ & $[4]_{X}[31]_{FS}[31]^{2}_{F}[22]_{S}$ \\
 
i  &  16    &  17   &&& \\

Config.  & $[4]_{X}[31]_{FS}[31]_{F}^{1}[31]_{S}$ & $[4]_{X}[31]_{FS}[31]_{F}^{2}[31]_{S}$  & & & \\

\hline \hline

\end{tabular}
\label{con} 
\end{table}
}

\newpage

{\squeezetable
\begin{table}[t]
\caption{\footnotesize 
Transition coupling ($T_i$) and energy due to the hyperfin interaction 
and mass difference between strange and light quark ($\Delta E_i$) 
for the five-quark configurations with light $q\bar{q}$ pairs and $L_{4q}=1$ in the octet baryons. 
The first row for each configuration with number i is the coupling $T_i$, followed by
rows for the energy $\Delta E_i$.}

\begin{tabular}{lccccc}
\hline 
\hline
                           i             &&   N  & $\Lambda $  &   $\Sigma$  & $\Xi$ \\
\hline
1
&&   $\frac{\sqrt{3}}{9}$ &    $\frac{\sqrt{2}}{6}$ & $\frac{\sqrt{6}}{18}$ &  $\frac{\sqrt{2}}{6}$    \\
%
&&   $-\frac{1}{3}(56P_{0}^{\pi}+28P_{1}^{\pi})$ &    $\delta m-\frac{1}{3}(28P_{0}^{\pi}+14P_{1}^{\pi}$
        & $\delta m-\frac{1}{3}(28P_{0}^{\pi}+14P_{1}^{\pi}$ 
        &  $2\delta m-\frac{1}{9}(8P_{0}^{\pi}+4P_{1}^{\pi}+152P_{0}^{K}$ \\
        &&   &   $+28P_{0}^{K}+14P_{1}^{K})$  & $+28P_{0}^{K}+14P_{1}^{K})$ 
        &  $+76P_{1}^{K}+8P_{0}^{s\bar{s}}+4P_{1}^{s\bar{s}})$ \\
2
&&    $\frac{\sqrt{6}}{9}$ &     $0$ & $\frac{4\sqrt{3}}{27}$ &  $\frac{\sqrt{3}}{9}$ \\
%
&&    $-\frac{1}{9}(128P_{0}^{\pi}+64P_{1}^{\pi})$ &     $-$ &  $\delta m-\frac{1}{9}(24P_{0}^{\pi}+12P_{1}^{\pi}$       
              &  $2\delta m-\frac{1}{9}(8P_{0}^{\pi}+4P_{1}^{\pi}+112P_{0}^{K}$  \\
&&                               &     $-$  &  $+104P_{0}^{K}+52P_{1}^{K})$       
              &  $+56P_{1}^{K}+8P_{0}^{s\bar{s}}+4P_{1}^{s\bar{s}})$ \\
3
&&    $0$ & $\frac{\sqrt{2}}{6}$ & $\frac{\sqrt{6}}{54}$ &  $\frac{\sqrt{2}}{18}$  \\
%
&&    $-$ &  $\delta m-\frac{1}{9}(84P_{0}^{\pi}+42P_{1}^{\pi}$
          &  $\delta m-\frac{1}{9}(84P_{0}^{\pi}+42P_{1}^{\pi}$       
          &  $2\delta m-\frac{1}{9}(48P_{0}^{\pi}+24P_{1}^{\pi}+72P_{0}^{K}$    \\

&&     &  $+44P_{0}^{K}+22P_{1}^{K})$   &  $+44P_{0}^{K}+22P_{1}^{K})$       
              &  $+36P_{1}^{K}+8P_{0}^{s\bar{s}}+4P_{1}^{s\bar{s}})$  \\
4
&& $0$& $\frac{\sqrt{2}}{18}$   & $\frac{\sqrt{6}}{18}$ & $\frac{\sqrt{6}}{18}$  \\ 
                        %
&& $-$ &   $\delta m-\frac{1}{36}(193P_{0}^{\pi}+31P_{1}^{\pi}$
& $\delta m-\frac{1}{36}(193P_{0}^{\pi}+31P_{1}^{\pi}$ & 
$2\delta m-\frac{1}{18}(50P_{0}^{\pi}+14P_{1}^{\pi}+183P_{0}^{K}$  \\

&&  &  $+283P_{0}^{K}+69P_{1}^{K})$ & $+283P_{0}^{K}+69P_{1}^{K})$ & 
$+41P_{1}^{K}-5P_{0}^{s\bar s}+5P_{1}^{s\bar s})$  \\
5
&&  $0$&  $\frac{\sqrt{2}}{18}$ &  $\frac{\sqrt{6}}{18}$ &  $\frac{\sqrt{6}}{18}$  \\
%
&& $-$  
              &  $\delta m-\frac{1}{18}(91P_{0}^{\pi}+17P_{1}^{\pi}$ 
              & $\delta m-\frac{1}{18}(91P_{0}^{\pi}+17P_{1}^{\pi}$ & 
$2\delta m-\frac{1}{9}(20P_{0}^{\pi}+8P_{1}^{\pi}+73P_{0}^{K}$ \\
&&  &  $+113P_{0}^{K}+19P_{1}^{K})$ & $+113P_{1}^{K}+19P_{1}^{K})$ & 
$+11P_{1}^{K}+9P_{0}^{s\bar s}+3P_{1}^{s\bar s})$  \\
6
&& $\frac{\sqrt{2}}{9}$ & $\frac{\sqrt{3}}{9}$  & $\frac{1}{9}$  & $\frac{\sqrt{3}}{9}$  \\
%
&&   $-\frac{1}{18}(158P_{0}^{\pi}+10P_{1}^{\pi})$ 
    &    $\delta m-\frac{1}{18}(79P_{0}^{\pi}+5P_{1}^{\pi}$  
     & $\delta m-\frac{1}{18}(79P_{0}^{\pi}+5P_{1}^{\pi}$ 
        &  $2\delta m-\frac{1}{27}(13P_{0}^{\pi}-P_{1}^{\pi}+211P_{0}^{K}$ \\
        &&  &   $+79P_{0}^{K}+5P_{1}^{K})$   & $+79P_{0}^{K}+5P_{1}^{K})$ 
        &  $+17P_{1}^{K}+13P_{0}^{s\bar{s}}-P_{1}^{s\bar{s}})$ \\
7
&&   $\frac{\sqrt{6}}{9}$&  $0$   &  $\frac{4\sqrt{3}}{27}$&  $\frac{\sqrt{3}}{9}$  \\
%
&&   $-\frac{1}{9}(73P_{0}^{\pi}-P_{1}^{\pi})$     &    $-$  
        & $\delta m-\frac{1}{9}(-5P_{0}^{\pi}-15P_{1}^{\pi}$ 
        &  $2\delta m-\frac{1}{9}(-4P_{1}^{\pi}+73P_{0}^{K}$ \\
        &&   & & $+78P_{0}^{K}+14P_{1}^{K})$ 
        &  $+7P_{1}^{K}-4P_{1}^{s\bar{s}})$ \\
8
&&  $0$ &  $\frac{\sqrt{2}}{6}$  &  $\frac{\sqrt{6}}{54}$ &  $\frac{\sqrt{2}}{18}$  \\
%
&&    $-$ &   $\delta m-\frac{1}{36}(229P_{0}^{\pi}+27P_{1}^{\pi}$ 
              &  $\delta m-\frac{1}{36}(229P_{0}^{\pi}+27P_{1}^{\pi}$       
              &  $2\delta m-\frac{1}{18}(78P_{0}^{\pi}+18P_{1}^{\pi}+73P_{0}^{K}$   \\
&&     &  $+63P_{0}^{K}-31P_{1}^{K})$   &  $+63P_{0}^{K}-31P_{1}^{K})$       
              &  $-9P_{1}^{K}-5P_{0}^{s\bar{s}}-5P_{1}^{s\bar{s}})$  \\
9
&& $-\frac{\sqrt{2}}{9}$ &  $0$  &  $-\frac{4}{27}$ &  $-\frac{1}{9}$  \\
%
&&    $-\frac{1}{27}(148P_{0}^{\pi}-4P_{1}^{\pi})$ &     $-$ 
              &  $\delta m-\frac{1}{81}(71P_{0}^{\pi}-59P_{1}^{\pi}$       
              &  $2\delta m-\frac{1}{81}(29P_{0}^{\pi}-17P_{1}^{\pi}+134P_{0}^{K}$ \\
&&  &       &  $+373P_{0}^{K}+47P_{1}^{K})$
&  $+10P_{1}^{K}+13P_{0}^{s\bar{s}}-25P_{1}^{s\bar{s}})$  \\
10
&&  $0$ &  $-\frac{\sqrt{6}}{18}$ &  $-\frac{\sqrt{2}}{54}$ &  $-\frac{\sqrt{6}}{54}$  \\
%
&&    $-$ &  $\delta m-\frac{1}{162}(595P_{0}^{\pi}+41P_{1}^{\pi}$   
&  $\delta m-\frac{1}{162}(595P_{0}^{\pi}+41P_{1}^{\pi}$       
              &  $2\delta m-\frac{1}{81}(180P_{0}^{\pi}+36P_{1}^{\pi}+235P_{0}^{K}$  \\
&&     &  $+293P_{0}^{K}-65P_{1}^{K})$    &  $+293P_{0}^{K}-65P_{1}^{K})$       
              &  $-31P_{1}^{K}+29P_{0}^{s\bar{s}}-17P_{1}^{s\bar{s}})$  \\
\hline
\hline
\end{tabular}
\label{eng1q}
\end{table}
}

\newpage

{\squeezetable
\begin{table}[t]
\caption{\footnotesize $T_i$ and $\Delta E_i$ for the five-quark configurations with 
$s\bar{s}$ pairs and $L_{4q}=1$ in the octet baryons. Conventions are the same as in
Table~\ref{eng1q}.}

\begin{tabular}{lccccc}
\hline 
\hline
             i                           &&   N  & $\Lambda $  &   $\Sigma$  & $\Xi$ \\
\hline

\hline
%
1
&&  $\frac{\sqrt{6}}{18}$ &  $0$ & $\frac{\sqrt{3}}{9}$ &  $0$ \\
%
&&  $2\delta m-\frac{1}{3}(28P_{0}^{\pi}+14P_{1}^{\pi}$  &    $-$ 
&  $3\delta m-\frac{1}{9}(8P_{0}^{\pi}+4P_{1}^{\pi}+152P_{0}^{K}$ & $-$   \\
        &&  $+28P_{0}^{K}+14P_{1}^{K})$  &  
        &  $+76P_{1}^{K}+8P_{0}^{s\bar{s}}+4P_{1}^{s\bar{s}})$
        &   \\
2
&& $0$ & $0$ & $\frac{\sqrt{2}}{9}$ &  $\frac{2}{9}$   \\
%
&&    $-$ &     $-$ &  $3\delta m-\frac{1}{9}(8P_{0}^{\pi}+4P_{1}^{\pi}+112P_{0}^{K}$       
              &  $4\delta m-\frac{1}{9}(104P_{0}^{K}+52P_{1}^{K}$ \\
&&      &  &  $+56P_{1}^{K}+8P_{0}^{s\bar{s}}+4P_{1}^{s\bar{s}})$      
              & $+24P_{0}^{s\bar s}+12P_{1}^{s\bar s})$  \\
3
&& $\frac{\sqrt{6}}{18}$ & $\frac{\sqrt{3}}{9}$   &  $0$  &  $0$  \\
%
&&    $2\delta m-\frac{1}{9}(84P_{0}^{\pi}+42P_{1}^{\pi}$ &     
          $3\delta m-\frac{1}{9}(48P_{0}^{\pi}+24P_{1}^{\pi}+72P_{0}^{K}$
& $-$  &  $-$   \\
&&    $+44P_{0}^{K}+22P_{1}^{K})$ 
& $+36P_{1}^{K}+8P_{0}^{s\bar{s}}+4P_{1}^{s\bar{s}})$       
              &   &   \\
4
&&  $\frac{\sqrt{6}}{18}$ &   $\frac{1}{9}$&   $0$ &   $0$   \\
%
 &&    $2\delta m-\frac{1}{36}(193P_{0}^{\pi}+31P_{1}^{\pi}$ &     
          $3\delta m-\frac{1}{18}(50P_{0}^{\pi}+14P_{1}^{\pi}+183P_{0}^{K}$
 & $-$  &  $-$   \\

&&    $+283P_{0}^{K}+69P_{1}^{K})$  
&  $+41P_{1}^{K}-5P_{0}^{s\bar{s}}+5P_{1}^{s\bar{s}})$  & &    \\
5
&& $\frac{\sqrt{6}}{18}$ &  $\frac{1}{9}$&  $0$ &  $0$  \\
%
 &&    $2\delta m-\frac{1}{18}(91P_{0}^{\pi}+17P_{1}^{\pi}$ &     
          $3\delta m-\frac{1}{9}(20P_{0}^{\pi}+8P_{1}^{\pi}+73P_{0}^{K}$& $-$        
              &  $-$   \\

&&    $+113P_{0}^{K}+19P_{1}^{K})$   
&  $+11P_{1}^{K}+9P_{0}^{s\bar{s}}+3P_{1}^{s\bar{s}})$ & & \\
6
&& $\frac{1}{9}$   &  $0$ & $\frac{\sqrt{2}}{9}$ &  $0$  \\
%
&&  $2\delta m-\frac{1}{18}(79P_{0}^{\pi}+5P_{1}^{\pi}$ &    $-$ 
&  $3\delta m-\frac{1}{27}(13P_{0}^{\pi}-P_{1}^{\pi}+211P_{0}^{K}$ &    $-$ \\

        &&  $+79P_{0}^{K}+5P_{1}^{K})$ &    
        & $+17P_{1}^{K}+13P_{0}^{s\bar{s}}-P_{1}^{s\bar{s}})$
        &       \\
7
&&  $0$ &  $0$ &  $\frac{\sqrt{2}}{9}$ &  $\frac{2}{9}$  \\
%
&&    $-$ &     $-$&  $3\delta m-\frac{1}{9}(-4P_{1}^{\pi}+73P_{0}^{K}$       
              &  $4\delta m-\frac{1}{9}(78P_{0}^{K}+14P_{1}^{K}$  \\

&&     &     &  $+7P_{1}^{K}-4P_{1}^{s\bar{s}})$      
              & $-5P_{0}^{s\bar s}-15P_{1}^{s\bar s})$     \\
8
&&  $\frac{\sqrt{6}}{18}$ &  $\frac{\sqrt{3}}{9}$ &  $0$ &  $0$ \\                                
%
  &&    $2\delta m-\frac{1}{36}(229P_{0}^{\pi}+27P_{1}^{\pi}$ &     
          $3\delta m-\frac{1}{18}(78P_{0}^{\pi}+18P_{1}^{\pi}+73P_{0}^{K}$
          & $-$  &  $-$   \\
&&    $+63P_{0}^{K}-31P_{1}^{K})$ &  $-9P_{1}^{K}-5P_{0}^{s\bar{s}}-5P_{1}^{s\bar{s}})$  
  &   &   \\
9
&&  $0$ &  $0$ &  $-\frac{\sqrt{6}}{27}$ &  $-\frac{2\sqrt{3}}{27}$ \\                                
%
&&    $-$& $-$ &  $3\delta m-\frac{1}{81}(29P_{0}^{\pi}-17P_{1}^{\pi}+134P_{0}^{K}$       
              &  $4\delta m-\frac{1}{81}(373P_{0}^{K}+47P_{1}^{K}$  \\

&&      &    &  $+10P_{1}^{K}+13P_{0}^{s\bar{s}}-25P_{1}^{s\bar{s}})$      
              & $+71P_{0}^{s\bar s}-59P_{1}^{s\bar s})$     \\
10
&&  $-\frac{\sqrt{2}}{18}$ &  $-\frac{1}{9}$  &  $0$ &  $0$  \\                                
%
  &&    $2\delta m-\frac{1}{162}(595P_{0}^{\pi}+41P_{1}^{\pi}$ &     
          $3\delta m-\frac{1}{81}(180P_{0}^{\pi}+36P_{1}^{\pi}+235P_{0}^{K}$ 
          & $-$   &  $-$  \\
ù
&&    $+293P_{0}^{K}-65P_{1}^{K})$   
&  $-31P_{1}^{K}+29P_{0}^{s\bar{s}}-17P_{1}^{s\bar{s}})$  &  &   \\
\hline
\hline
\end{tabular}
\label{eng1s}
\end{table}
}

\newpage

{\squeezetable
\begin{table}[t]
\caption{\footnotesize $T_i$ and $\Delta E_i$ for the five-quark configurations with 
$L_{4q}=0$ in the octet baryons. Upper and lower panels are for
the configurations with light $q\bar{q}$ and $s\bar{s}$ pairs, respectively. Conventions
are the same as in Table~\ref{eng1q}.}

\begin{tabular}{lccccc}
\hline \hline
    i                                    &&   N  & $\Lambda $  &   $\Sigma$  & $\Xi$ \\
\hline
11
&& $0$ & $-\frac{\sqrt{5}}{18}$ & $-\frac{\sqrt{15}}{18}$ & $-\frac{\sqrt{15}}{18}$  \\
%
&& $-$ & $\delta m-\frac{56}{9}P^{\pi}_{0}-\frac{88}{9}P^{K}_{0}$  
              & $\delta m-\frac{56}{9}P^{\pi}_{0}-\frac{88}{9}P^{K}_{0}$
              & $2\delta m-\frac{32}{9}P^{\pi}_{0}-\frac{112}{9}P^{K}_{0}$   \\
12
&&  $0$ &  $-\frac{\sqrt{5}}{18}$ &  $-\frac{\sqrt{15}}{18}$ & $-\frac{\sqrt{15}}{18}$ \\

&&  $-$  
              &  $\delta m-6P^{\pi}_{0}-\frac{22}{3}P^{K}_{0}$
              &  $\delta m-6P^{\pi}_{0}-\frac{22}{3}P^{K}_{0}$ 
              &  $2\delta m-\frac{8}{3}P^{\pi}_{0}-\frac{28}{3}P^{K}_{0}-\frac{4}{3}P^{s\bar{s}}_{0}$ \\
13
&& $-\frac{\sqrt{5}}{9}$ & $-\frac{\sqrt{30}}{18}$ & $-\frac{\sqrt{10}}{18}$ & $-\frac{\sqrt{30}}{18}$ \\

&& $-\frac{28}{3}P^{\pi}_{0}$  
              & $\delta m-\frac{14}{3}P^{\pi}_{0}-\frac{14}{3}P^{K}_{0}$
              & $\delta m-\frac{14}{3}P^{\pi}_{0}-\frac{14}{3}P^{K}_{0}$ 
              & $2\delta m-\frac{4}{9}P^{\pi}_{0}-\frac{76}{9}P^{K}_{0}-\frac{4}{9}P^{s\bar{s}}_{0}$ \\
14
&&   $-\frac{\sqrt{15}}{9}$ &  $0$ &  $-\frac{2\sqrt{30}}{27}$ &  $-\frac{\sqrt{30}}{18}$  \\

&&   $-8P^{\pi}_{0}$ &  $-$ 
              &  $\delta m+\frac{20}{9}P^{\pi}_{0}-\frac{92}{9}P^{K}_{0}$ 
              &  $2\delta m+\frac{4}{9}P^{\pi}_{0}-\frac{80}{9}P^{K}_{0}+\frac{4}{9}P^{s\bar{s}}_{0}$  \\
15
&&  $0$  &  $-\frac{\sqrt{5}}{6}$ &  $-\frac{\sqrt{15}}{54}$ &  $-\frac{\sqrt{5}}{18}$  \\
%
 &&  $-$  &  $\delta m-\frac{64}{9}P^{\pi}_{0}-\frac{8}{9}P^{K}_{0}$
               &  $\delta m-\frac{64}{9}P^{\pi}_{0}-\frac{8}{9}P^{K}_{0}$ 
              &  $2\delta m-\frac{16}{3}P^{\pi}_{0}-\frac{32}{9}P^{K}_{0}+\frac{8}{9}P^{s\bar{s}}_{0}$ \\
16
&&   $\frac{\sqrt{5}}{9}$ &  $0$  &  $\frac{2\sqrt{10}}{27}$ &  $\frac{\sqrt{10}}{18}$ \\
%
&&   $-\frac{16}{3}P^{\pi}_{0}$ &  $-$  &  $\delta m-\frac{4}{27}P^{\pi}_{0}-\frac{140}{27}P^{K}_{0}$ 
         &  $2\delta m-\frac{4}{27}P^{\pi}_{0}-\frac{16}{3}P^{K}_{0}+\frac{4}{27}P^{s\bar{s}}_{0}$ \\
17
&&  $0$ &  $\frac{\sqrt{15}}{18}$ &  $\frac{\sqrt{5}}{54}$ &  $\frac{\sqrt{15}}{54}$  \\

&&  $-$ &  $\delta m-\frac{106}{27}P^{\pi}_{0}-\frac{38}{27}P^{K}_{0}$
              &  $\delta m-\frac{106}{27}P^{\pi}_{0}-\frac{38}{27}P^{K}_{0}$ 
      &  $2\delta m-\frac{8}{3}P^{\pi}_{0}-\frac{68}{27}P^{K}_{0}-\frac{4}{27}P^{s\bar{s}}_{0}$  \\
\hline
11
&&  $-\frac{\sqrt{15}}{18}$ & $-\frac{\sqrt{10}}{18}$   &   $0$ &   $0$  \\
%
&&  $2\delta m-\frac{56}{9}P^{\pi}_{0}-\frac{88}{9}P^{K}_{0}$ 
              &   $3\delta m-\frac{32}{9}P^{\pi}_{0}-\frac{112}{9}P^{K}_{0}$  
              &   $-$ &   $-$       \\
12
&&  $-\frac{\sqrt{15}}{18}$ &   $-\frac{\sqrt{10}}{18}$  &   $0$ &   $0$   \\
%
&& $2\delta m-6P^{\pi}_{0}-\frac{22}{3}P^{K}_{0}$ 
              &  $3\delta m-\frac{8}{3}P^{\pi}_{0}-\frac{28}{3}P^{K}_{0}-\frac{4}{3}P^{s\bar{s}}_{0}$ 
              &  $-$ &  $-$ \\
13
&& $-\frac{\sqrt{10}}{18}$ &   $0$ & $-\frac{\sqrt{5}}{9}$ &  $0$ \\

&& $2\delta m-\frac{14}{3}P^{\pi}_{0}-\frac{14}{3}P^{K}_{0}$ &   $-$   
    & $3\delta m-\frac{4}{9}P^{\pi}_{0}-\frac{76}{9}P^{K}_{0}-\frac{4}{9}P^{s\bar{s}}_{0}$ &  $-$  \\

14
&&  $0$ &  $0$  &  $-\frac{\sqrt{5}}{9}$ &  $-\frac{\sqrt{10}}{9}$  \\                                
                          
&&  $-$ &  $-$ 
&  $3\delta m+\frac{4}{9}P^{\pi}_{0}-\frac{80}{9}P^{K}_{0}+\frac{4}{9}P^{s\bar{s}}_{0}$  
              &  $4\delta m-\frac{92}{9}P^{K}_{0}+\frac{20}{9}P^{s\bar{s}}_{0}$  \\
15
&&  $-\frac{\sqrt{15}}{18}$  &  $-\frac{\sqrt{30}}{18}$ &  $0$ &  $0$  \\
%
&&  $2\delta m-\frac{64}{9}P^{\pi}_{0}-\frac{8}{9}P^{K}_{0}$    
   &  $3\delta m-\frac{16}{3}P^{\pi}_{0}-\frac{32}{9}P^{K}_{0}+\frac{8}{9}P^{s\bar{s}}_{0}$ 
   & $-$  &  $-$ \\
16
&&  $0$ &  $0$ &  $\frac{\sqrt{15}}{27}$  &  $\frac{\sqrt{30}}{27}$  \\                                
%
&&  $-$  &  $-$  
     &  $3\delta m-\frac{4}{27}P^{\pi}_{0}-\frac{16}{3}P^{K}_{0}+\frac{4}{27}P^{s\bar{s}}_{0}$ 
              &  $4\delta m-\frac{140}{27}P^{K}_{0}-\frac{4}{27}P^{s\bar{s}}_{0}$ \\
17
&&  $\frac{\sqrt{5}}{18}$ &  $\frac{\sqrt{10}}{18}$ &  $0$ &  $0$  \\
%
&&  $2\delta m-\frac{106}{27}P^{\pi}_{0}-\frac{38}{27}P^{K}_{0}$   
   &  $3\delta m-\frac{8}{3}P^{\pi}_{0}-\frac{68}{27}P^{K}_{0}-\frac{4}{27}P^{s\bar{s}}_{0}$ 
              &  $-$  &  $-$ \\
\hline \hline
\end{tabular}
\label{eng0}
\end{table}
}

\newpage

{\squeezetable
\begin{table}[ht]
\caption{\footnotesize Predictions for probabilities of different five-quark components in the nucleon, 
$\Lambda$, $\Sigma$ and $\Xi$.
Reported values correspond to $\bar{d}-\bar{u}$ for the proton in the range 0.118 $\pm$ 0.012. 
Whenever the calculated probability varies in that range by less than 0.001, a single non vanishing 
value is given.
Upper and lower panels are for the configurations with $L_{4q}=1$ and $L_{4q}=0$, respectively. 
}
\begin{tabular}{ccccccc}
\hline \hline
  Configuration  &             &&   N    &  $\Lambda $ & $\Sigma$  & $\Xi$ \\
\hline
1

& $q\bar{q}$ &&  0.146 $\pm$ 0.015     & 0.114 $\pm$ 0.013   & 0.067 $\pm$ 0.008   & 0.082 $\pm$ 0.009 \\
                           
& $s\bar{s}$ &&  0.010 $\pm$ 0.001 & 0 & 0.020 $\pm$ 0.002 & 0     \\
                         
2

& $q\bar{q}$ &&    0.073 $\pm$ 0.007     & 0   & 0.055 $\pm$ 0.006  & 0.028 $\pm$ 0.003 \\

& $s\bar{s}$ &&  0  & 0        & 0.009 $\pm$ 0.001   & 0.016 $\pm$ 0.002      \\

3

& $q\bar{q}$ &&    0 & 0.052 $\pm$ 0.006   & 0.003 $\pm$ 0.001   &  0.006 $\pm$ 0.001 \\

& $s\bar{s}$ &&   0.006 $\pm$ 0.001  & 0.013 $\pm$ 0.002    & 0  & 0   \\

4

& $q\bar{q}$ && 0 & 0.003$\pm$ 0.001 & 0.010 $\pm$ 0.001  & 0.010 $\pm$ 0.002 \\                           

& $s\bar{s}$ && 0.004 $\pm$ 0.001  & 0.003 $\pm$ 0.001  & 0   & 0   \\

5

&  $q\bar{q}$ &&  0 & 0.002  & 0.008 $\pm$ 0.001 &  0.008 $\pm$ 0.001 \\

& $s\bar{s}$ &&   0.003 $\pm$ 0.001 & 0.002 $\pm$ 0.001   & 0  & 0   \\

6

& $q\bar{q}$ && 0.006 $\pm$ 0.001  & 0.010 $\pm$ 0.001 & 0.004 & 0.011 $\pm$ 0.002 \\

& $s\bar{s}$ &&   0.002 & 0   & 0.004 $\pm$ 0.001  & 0   \\

7

& $q\bar{q}$ && 0.016 $\pm$ 0.002 & 0 & 0.017 $\pm$ 0.002 &  0.010  $\pm$ 0.002 \\

& $s\bar{s}$ &&  0  & 0   & 0.004 $\pm$ 0.001  & 0.009 $\pm$ 0.002    \\
                          
8

& $q\bar{q}$ &&  0  & 0.015 $\pm$ 0.002 & 0.001  &  0.002  \\

& $s\bar{s}$ && 0.003 $\pm$ 0.001  & 0.006 $\pm$ 0.001   & 0   & 0  \\
                          
9

& $q\bar{q}$ &&  0.005 $\pm$ 0.001 & 0 &  0.005 $\pm$ 0.001  &  0.003 $\pm$ 0.001  \\

& $s\bar{s}$ &&   0  & 0   & 0.001   & 0.003 $\pm$ 0.001    \\
                          
10

& $q\bar{q}$ &&  0 &  0.004 $\pm$ 0.001  & 0  & 0.001 $\pm$ 0.001  \\
                          
& $s\bar{s}$ && 0.001  & 0.002  & 0  & 0 \\
                          
\hline


11

& $q\bar{q}$ && 0   &   0.005 $\pm$ 0.001 &   0.019 $\pm$ 0.002 & 0.018 $\pm$ 0.002 \\

& $s\bar{s}$ && 0.009 $\pm$ 0.001  & 0.006 $\pm$ 0.001  & 0   & 0 \\

12
 
& $q\bar{q}$ && 0  &  0.003 $\pm$ 0.001 &  0.007 $\pm$ 0.001 & 0.006 $\pm$ 0.001 \\

& $s\bar{s}$ && 0.008 $\pm$ 0.001  & 0.006 $\pm$ 0.001   & 0  & 0  \\

13

& $q\bar{q}$ && 0.015 $\pm$ 0.002  &  0.025 $\pm$ 0.003 &  0.010 $\pm$ 0.001 & 0.028 $\pm$ 0.003 \\

& $s\bar{s}$ && 0.004 $\pm$ 0.001  & 0  & 0.011 $\pm$ 0.001  & 0     \\

14

& $q\bar{q}$ &&  0.041 $\pm$ 0.004  & 0  & 0.043 $\pm$ 0.005 &  0.026 $\pm$ 0.003 \\
                          
& $s\bar{s}$ &&  0  & 0  & 0.010 $\pm$ 0.002  & 0.022 $\pm$ 0.003  \\

15

& $q\bar{q}$ &&  0  &  0.036 $\pm$ 0.004 &  0.002 $\pm$ 0.001 & 0.005 $\pm$ 0.001 \\                                
                          
& $s\bar{s}$ &&  0.007 $\pm$ 0.001   & 0.015 $\pm$ 0.002     & 0  & 0 \\

16

& $q\bar{q}$ &&  0.012 $\pm$ 0.001  & 0  & 0.013 $\pm$ 0.002  &  0.008 $\pm$ 0.001 \\                                
                          
& $s\bar{s}$ &&  0   & 0  & 0.003 $\pm$ 0.001   & 0.007 $\pm$ 0.001    \\

17

& $q\bar{q}$ && 0  &  0.010 $\pm$ 0.001  &  0  &  0.001 $\pm$ 0.001 \\                                
                          
& $s\bar{s}$ &&  0.002   & 0.004 $\pm$ 0.001   &  0   & 0   \\

\hline\hline
                              
\end{tabular}
\label{nump}
\end{table}
}
%

{\squeezetable
\begin{table}[ht]
\caption{\footnotesize Predictions for the sea content of the octet baryons. 
The experimental $\bar{d}-\bar{u}$ flavor asymmetry value for the proton  
$A_p$={\it 0.118  $\pm$ 0.012}, is used as input. 
(Notice that $A_n=-A_p$, $A_{\Sigma^-}=-A_{\Sigma^+}$ and $A_{\Xi^-}=-A_{\Xi^\circ}$).}

\begin{tabular}{lcccccccc}
\hline \hline
  Baryon   &   $\bar{u}$      &   $\bar{d}$   & $\bar{s}$  & $\bar{d}-\bar{u}$ & $\bar{u}+\bar{d}$  & $\bar{u}+\bar{d}+\bar{s}$ & Approach & Reference\\
\hline
p              &   0.098 $\pm$ 0.010  &  0.216 $\pm$ 0.022  & 0.057 $\pm$ 0.006  & {\it 0.118  $\pm$ 0.012} & 0.314 $\pm$ 0.032  & 0.371 $\pm$ 0.038 & E-$\chi$CQM & Present work \\
               &   0.228  &  0.358 & -  & 0.130 &  0.586  & - & $\chi$CQM & Shao {\it et al.}~\cite{arXiv:1002.4747}\\
               &   0.033 - 0.325  &  0.163 - 0.455 & -  & 0.130 & 0.196 - 0.880   & - & MC & Shao {\it et al.}~\cite{arXiv:1002.4747}\\
              &   0.176          &  0.294         & -         &  {\it 0.118} &  0.470  & - & G-BHPS & Chang-Peng~\cite{arXiv:1102.5631} \\
              &   0.122          &  0.240         & 0.024     &  {\it 0.118} &  0.362  & 0.386 & G-BHPS & Chang-Peng~\cite{arXiv:1105.2381} \\
              &   0.162          &  0.280         & 0.029     & {\it 0.118}  &  0.442  & 0.379 & G-BHPS & Chang-Peng~\cite{arXiv:1105.2381} \\
              &             &           &      & 0.151  &    &  & UqQM & Santopinto-Bijker~\cite{885054} \\
              [5pt]
$\Lambda$     &  0.139  $\pm$ 0.015 & 0.139 $\pm$ 0.015 & 0.057 $\pm$ 0.006 & 0 & 0.279 $\pm$ 0.030 & 0.336 $\pm$ 0.036 & E-$\chi$CQM & Present work \\
               &   0.195  &  0.195 & -  & 0 &  0.390  & - & $\chi$CQM & Shao {\it et al.}~\cite{arXiv:1002.4747}\\
               &   0.098 - 0.390  &  0.098 - 0.390 & -  & 0 & 0.196 - 0.780   & - & MC& Shao {\it et al.}~\cite{arXiv:1002.4747}\\
              [5pt]
$\Sigma^{+}$  &   0.100 $\pm$ 0.011 &  0.163 $\pm$ 0.018& 0.063 $\pm$ 0.007 & 0.063 $\pm$ 0.007 &  0.263 $\pm$ 0.029 & 0.326 $\pm$ 0.036& E-$\chi$CQM & Present work \\
               &   0.065  &  0.325 & -  & 0.260 &  0.390  & - & $\chi$CQM & Shao {\it et al.}~\cite{arXiv:1002.4747}\\
               &   0.049 - 0.164  &  0.341 - 0.839 & -  & 0.293 - 0.675 & 0.390 - 1.001   & - & MC & Shao {\it et al.}~\cite{arXiv:1002.4747}\\
              &             &           &      & 0.126  &    &  & UqQM & Santopinto-Bijker~\cite{885054} \\
              [5pt]
$\Sigma^{0}$  &  0.132 $\pm$ 0.015 & 0.132 $\pm$ 0.015 & 0.063 $\pm$ 0.007  & 0 &  0.263 $\pm$ 0.029  & 0.326 $\pm$ 0.036& E-$\chi$CQM & Present work \\
               &  0.195  &  0.195 & -  & 0 &  0.390  & - & $\chi$CQM & Shao {\it et al.}~\cite{arXiv:1002.4747}\\
               &   0.195 - 0.501  &  0.195 - 0.501 & -  & 0 & 0.390 - 1.001   & - & MC & Shao {\it et al.}~\cite{arXiv:1002.4747}\\
              [5pt]
$\Xi^{0}$     &  0.131 $\pm$ 0.015 & 0.121 $\pm$ 0.014 & 0.057 $\pm$ 0.006 & -0.009 $\pm$ 0.001 & 0.252 $\pm$ 0.028 & 0.309 $\pm$ 0.035 & E-$\chi$CQM & Present work \\
               &   0.033  &  0.163 & -  & 0.130 &  0.196  & - & $\chi$CQM & Shao {\it et al.}~\cite{arXiv:1002.4747}\\
               &   0.033 - 0.130  & 0.163 - 0.650 & -  & 0.130 - 0.520 & 0.199 - 0.780   & - & MC & Shao {\it et al.}~\cite{arXiv:1002.4747}\\
              &             &           &      & -0.001  &    &  & UqQM & Santopinto-Bijker~\cite{885054} \\
\hline\hline
\end{tabular}
\label{sea}
\end{table}
}

\begin{thebibliography}{99}
%

%
\bibitem{Thomas:1983fh}
  A.~W.~Thomas,
  Phys.\ Lett.\  B {\bf 126}, 97 (1983).

\bibitem{Arneodo:1994sh}
  M.~Arneodo {\it et al.}  [New Muon Collaboration],
  Phys.\ Rev.\  D {\bf 50}, 1 (1994).

\bibitem{Baldit:1994jk}
  A.~Baldit {\it et al.}  [NA51 Collaboration],
  Phys.\ Lett.\  B {\bf 332}, 244 (1994).

\bibitem{Hawker:1998ty}
  E.~A.~Hawker {\it et al.}  [FNAL E866/NuSea Collaboration],
  Phys.\ Rev.\ Lett.\  {\bf 80}, 3715 (1998).
  
\bibitem{Peng:1998pa}
  J.~C.~Peng {\it et al.}  [E866/NuSea Collaboration],
  Phys.\ Rev.\  D {\bf 58}, 092004 (1998).
    
\bibitem{Towell:2001nh}
  R.~S.~Towell {\it et al.}  [FNAL E866/NuSea Collaboration],
  Phys.\ Rev.\  D {\bf 64}, 052002 (2001).

\bibitem{Ackerstaff:1998sr}
  K.~Ackerstaff {\it et al.}  [HERMES Collaboration],
  Phys.\ Rev.\ Lett.\  {\bf 81}, 5519 (1998).

\bibitem{Gottfried:1967kk}
  K.~Gottfried,
  Phys.\ Rev.\ Lett.\  {\bf 18}, 1174 (1967).
  
\bibitem{Field:1976ve}
  R.~D.~Field and R.~P.~Feynman,
  Phys.\ Rev.\  D {\bf 15}, 2590 (1977).

\bibitem{Garvey:2001yq}
  G.~T.~Garvey and J.~C.~Peng,
  Prog.\ Part.\ Nucl.\ Phys.\  {\bf 47}, 203 (2001).

\bibitem{Speth:1996pz}
  J.~Speth and A.~W.~Thomas,
  Adv.\ Nucl.\ Phys.\  {\bf 24}, 83 (1997).

\bibitem{Kumano:1997cy}
  S.~Kumano,
  Phys.\ Rept.\  {\bf 303}, 183 (1998).

\bibitem{Vogt:2000sk}
  R.~Vogt,
  Prog.\ Part.\ Nucl.\ Phys.\  {\bf 45}, S105 (2000).
  
\bibitem{Peng:2003zm}
  J.~C.~Peng,
  Eur.\ Phys.\ J.\  A {\bf 18}, 395 (2003)

\bibitem{Diakonov:1995zi}
  D.~Diakonov,
  Prog.\ Part.\ Nucl.\ Phys.\  {\bf 36}, 1 (1996)

%
%
\bibitem{Henley:1990kw}
  E.~M.~Henley and G.~A.~Miller,
  Phys.\ Lett.\  B {\bf 251}, 453 (1990).
  
\bibitem{Alberg:2012wr}
  M.~Alberg and G.~A.~Miller,
  arXiv:1201.4184 [nucl-th].
%
%
\bibitem{Brodsky:1996hc}
  S.~J.~Brodsky and B.~Q.~Ma,
  Phys.\ Lett.\  B {\bf 381}, 317 (1996).

\bibitem{Cao:1999da}
  F.~G.~Cao and A.~I.~Signal,
  Phys.\ Rev.\  D {\bf 60}, 074021 (1999).

\bibitem{Cao:1999jc}
  F.~G.~Cao and A.~I.~Signal,
  Phys.\ Lett.\  B {\bf 474}, 138 (2000).
%
%
\bibitem{Nikolaev:1998se}
  N.~N.~Nikolaev, W.~Schafer, A.~Szczurek and J.~Speth,
  Phys.\ Rev.\  D {\bf 60}, 014004 (1999).
%
%
\bibitem{Carvalho:1999he}
  F.~Carvalho, F.~O.~Duraes, F.~S.~Navarra, M.~Nielsen and F.~M.~Steffens,
  Eur.\ Phys.\ J.\  C {\bf 18}, 127 (2000).

\bibitem{Huang:2007dy}
  F.~Huang, F.~G.~Cao and B.~Q.~Ma,
  Phys.\ Rev.\  D {\bf 76}, 114016 (2007).

\bibitem{Chen:2009xy}
  H.~Chen, F.~G.~Cao and A.~I.~Signal,
  J.\ Phys.\ G {\bf 37}, 105006 (2010).
%
%
\bibitem{Alwall:2005xd}
  J.~Alwall and G.~Ingelman,
  Phys.\ Rev.\  D {\bf 71}, 094015 (2005).


\bibitem{Brodsky:1988ip}
  S.~J.~Brodsky, J.~R.~Ellis and M.~Karliner,
  Phys.\ Lett.\  B {\bf 206}, 309 (1988).

\bibitem{Diakonov:1997mm}
  D.~Diakonov, V.~Petrov and M.~V.~Polyakov,
  Z.\ Phys.\  A {\bf 359}, 305 (1997).

\bibitem{Wakamatsu:1997en}
  M.~Wakamatsu and T.~Kubota,
  Phys.\ Rev.\  D {\bf 57}, 5755 (1998).

\bibitem{Dressler:1998zi}
  B.~Dressler, K.~Goeke, P.~V.~Pobylitsa, M.~V.~Polyakov, T.~Watabe and C.~Weiss,
  arXiv:hep-ph/9809487.

\bibitem{Pobylitsa:1998tk}
  P.~V.~Pobylitsa, M.~V.~Polyakov, K.~Goeke, T.~Watabe and C.~Weiss,
  Phys.\ Rev.\  D {\bf 59}, 034024 (1999).

\bibitem{615110} 
  M.~Wakamatsu,
  Phys.\ Rev.\ D\ {\bf 67}, 034005  (2003).  




\bibitem{Dahiya:2003ms}
  H.~Dahiya and M.~Gupta,
  Phys.\ Rev.\  D {\bf 67}, 074001 (2003).
  
\bibitem{Shu:2007xb}
  Z.~Shu, X.~L.~Chen and W.~Z.~Deng,
  Phys.\ Rev.\  D {\bf 75}, 094018 (2007).

\bibitem{885054} 
  E.~Santopinto and R.~Bijker,
  Phys.\ Rev.\ C\ {\bf 82}, 062202  (2010).  

\bibitem{arXiv:1002.4747} 
  L.~Shao, Y.~-J.~Zhang and B.~-Q.~Ma,
  Phys.\ Lett.\ B\ {\bf 686}, 136  (2010).

\bibitem{Sharma:2010sm}
  N.~Sharma and H.~Dahiya,
  Phys.\ Rev.\  D {\bf 81}, 114003 (2010).

\bibitem{Dahiya:2010ht}
  H.~Dahiya and N.~Sharma,
  Mod.\ Phys.\ Lett.\  A {\bf 26}, 279 (2011).

\bibitem{arXiv:1101.3378} 
  H.~Song, X.~Zhang and B.~-Q.~Ma,
  Eur.\ Phys.\ J.\ C\ {\bf 71}, 1542  (2011).
  
\bibitem{Yuan:2012wz} 
  S.~G.~Yuan, K.~W.~Wei, J.~He, H.~S.~Xu and B.~S.~Zou,
  arXiv:1201.0807 [nucl-th].  

 
\bibitem{Brodsky:1980pb}
  S.~J.~Brodsky, P.~Hoyer, C.~Peterson and N.~Sakai,
  Phys.\ Lett.\  B {\bf 93}, 451 (1980).

\bibitem{Brodsky:1981se}
  S.~J.~Brodsky, C.~Peterson and N.~Sakai,
  Phys.\ Rev.\  D {\bf 23}, 2745 (1981).

\bibitem{arXiv:1102.5631} 
  W.~-C.~Chang and J.~-C.~Peng,
  Phys.\ Rev.\ Lett.\ \ {\bf 106}, 252002  (2011).  

\bibitem{arXiv:1105.2381} 
  W.~-C.~Chang and J.~-C.~Peng,
  Phys.\ Lett.\ B\ {\bf 704}, 197  (2011).


 
\bibitem{Young:2009ps}
  R.~D.~Young and A.~W.~Thomas,
  Nucl.\ Phys.\  A {\bf 844}, 266C (2010).

\bibitem{Takeda:2010cw}
  K.~Takeda, S.~Aoki, S.~Hashimoto, T.~Kaneko, J.~Noaki and T.~Onogi  [JLQCD
                  collaboration],
  Phys.\ Rev.\  D {\bf 83}, 114506 (2011).

\bibitem{Bali:2011rs}
  G.~S.~Bali {\it et al.}  [QCDSF Collaboration and QCDSF Collaboration and
                  QCDSF Collaboration an],
  arXiv:1112.0024 [hep-lat].


\bibitem{Trevisan:2008zz}
  L.~A.~Trevisan, C.~Mirez, T.~Frederico and L.~Tomio,
  Eur.\ Phys.\ J.\  C {\bf 56}, 221 (2008).

\bibitem{Zhang:2010ac}
  B.~Zhang and Y.~J.~Zhang,
  Phys.\ Rev.\  D {\bf 82}, 074021 (2010).


\bibitem{Zou:2005xy}
  B.~S.~Zou and D.~O.~Riska,
  Phys.\ Rev.\ Lett.\  {\bf 95}, 072001 (2005).

\bibitem{Li:2005jn}
  Q.~B.~Li and D.~O.~Riska,
  Phys.\ Rev.\  C {\bf 73}, 035201 (2006).

\bibitem{An:2005cj}
  C.~S.~An, D.~O.~Riska and B.~S.~Zou
  Phys.\ Rev.\  C {\bf 73}, 035207 (2006).

\bibitem{JuliaDiaz:2006av}
  B.~Julia-Diaz and D.~O.~Riska,
  Nucl.\ Phys.\  A {\bf 780}, 175 (2006).

\bibitem{Li:2005jb}
  Q.~B.~Li and D.~O.~Riska,
  Nucl.\ Phys.\  A {\bf 766}, 172 (2006).

\bibitem{Li:2006nm}
  Q.~B.~Li and D.~O.~Riska,
  Phys.\ Rev.\  C {\bf 74}, 015202 (2006).

\bibitem{An:2008xk}
  C.~S.~An and B.~S.~Zou,
  Eur.\ Phys.\ J.\  A {\bf 39}, 195 (2009).

\bibitem{An:2008tz}
  C.~S.~An and D.~O.~Riska,
  Eur.\ Phys.\ J.\  A {\bf 37}, 263 (2008).

\bibitem{An:2010wb}
  C.~S.~An, B.~Saghai, S.~G.~Yuan and J.~He,
  Phys.\ Rev.\  C {\bf 81}, 045203 (2010).

\bibitem{Zou:2009zz}
  B.~S.~Zou,
  Nucl.\ Phys.\  A {\bf 827}, 333C (2009).

\bibitem{Zou:2010tc}
  B.~S.~Zou,
  Nucl.\ Phys.\  A {\bf 835}, 199 (2010).

\bibitem{An:2011sb}
  C.~An and B.~Saghai,
  Phys.\ Rev.\  C {\bf 84}, 045204 (2011).



\bibitem{chen}J. Q. Chen, Group Representation Theory for Physicists,
 World Scientific, Singapore (1989).

\bibitem{Glozman:1995fu}
  L.~Y.~Glozman and D.~O.~Riska,
  Phys.\ Rept.\  {\bf 268}, 263 (1996).
  
  
\bibitem{Bazarko:1994tt}
  A.~O.~Bazarko {\it et al.}  [CCFR Collaboration],
  Z.\ Phys.\  C {\bf 65}, 189 (1995).
  
\bibitem{Conrad:1997ne}
  J.~M.~Conrad, M.~H.~Shaevitz and T.~Bolton,
  Rev.\ Mod.\ Phys.\  {\bf 70}, 1341 (1998).


\bibitem{arXiv:0710.5032} 
  F.~X.~Wei and B.~S.~Zou,
  Phys.\ Lett.\ B\ {\bf 660}, 501  (2008).

\bibitem{Zeller:2001hh}
  G.~P.~Zeller {\it et al.}  [NuTeV Collaboration],
  Phys.\ Rev.\ Lett.\  {\bf 88}, 091802 (2002)
  [Erratum-ibid.\  {\bf 90}, 239902 (2003)].

\bibitem{Londergan:2007zza}
  J.~T.~Londergan,
  Eur.\ Phys.\ J.\  A {\bf 32}, 415 (2007).

\bibitem{Avakian:2012ca}
  H.~Avakian {\it et al.},
  arXiv:1202.1910 [hep-ex].
  
%

\bibitem{Yang:2009hh}
  R.~Yang, J.~C.~Peng and M.~Grosse-Perdekamp,
  Phys.\ Lett.\  B {\bf 680}, 231 (2009).

\bibitem{Brodsky:2012vg}
  S.~J.~Brodsky, F.~Fleuret, C.~Hadjidakis and J.~P.~Lansberg,
  arXiv:1202.6585 [hep-ph].


\bibitem{Adams:2009kp}
  T.~Adams {\it et al.}  [NuSOnG Collaboration],
  Int.\ J.\ Mod.\ Phys.\  A {\bf 25}, 909 (2010).

\bibitem{Dinter:2011zz}
  S.~Dinter, V.~Drach and K.~Jansen,
  Int.\ J.\ Mod.\ Phys.\ Proc.\ Suppl.\  E {\bf 20}, 110 (2011).

\bibitem{Giedt:2009mr}
  J.~Giedt, A.~W.~Thomas and R.~D.~Young,
  Phys.\ Rev.\ Lett.\  {\bf 103}, 201802 (2009).

\end{thebibliography}
\end{document}